\pdfoutput=1

\documentclass[prl,aps,superscriptaddress,amsmath,amssymb,nofootinbib,twocolumn]{revtex4-2}

\usepackage{braket,verbatim,bm,bbold,amsmath,amssymb,epsfig,float,color}
\usepackage[colorlinks=true,linkcolor=blueG, citecolor=redG, urlcolor=magentaG, bookmarks]{hyperref}
\usepackage{color}
\usepackage[dvipsnames]{xcolor}
\usepackage{enumitem}   
\usepackage{soul}
\usepackage{amsmath,amsfonts,amssymb,stackengine,graphicx}
\setstackgap{S}{-1.5pt}

\usepackage{tikz}
\usepackage{etoolbox}

\newcommand{\nc}{\newcommand}
\nc{\ir}{\mathrm{i}}
\nc{\dd}{\mathrm{d}} 
\nc{\eE}{\mathsf{e}}
\nc{\Tr}{\text{Tr}}
\nc{\id}{\mathbb{I}}
\nc{\I}{\mathcal{I}}
\nc{\F}{\mathcal{F}}
\nc{\M}{\mathcal{M}}
\nc{\Ll}{\mathcal{L}}
\nc{\gap}{\gamma_+}
\nc{\gam}{\gamma_-}
\nc{\xilatt}{\xi_{\textrm{latt}}}
\nc{\A}{\mathcal{A}}
\nc{\B}{\mathcal{B}}

\definecolor{blueG}{RGB}{51, 102, 204}
\definecolor{magentaG}{RGB}{214.2, 40.8, 132.6}
\definecolor{redG}{RGB}{229.5, 51., 102.}

\begin{document}

%%%%%%%%%%%%%%%%%%%%%%%%%%%%%%%%%%%%%%%%%%%%%%%%%%%%%%%%%%%%%%%%%%% 
\title{Universal scaling of spatially extended zero modes in inhomogeneous SSH chains}
%%%%%%%%%%%%%%%%%%%%%%%%%%%%%%%%%%%%%%%%%%%%%%%%%%%%%%%%%%%%%%%%%%%

\author{Gilles Parez}
\email{parez@lapth.cnrs.fr}
\affiliation{Laboratoire d'Annecy de Physique Th\'eorique (LAPTh), CNRS, Universit\'e Savoie Mont-Blanc, 74940 Annecy, France}

\author{Nicolas Cramp\'e}
\affiliation{Institut Denis-Poisson, CNRS, Université de Tours--Université d’Orléans,
37200 Tours, France}

\author{Quentin Labriet}
\affiliation{Centre de Recherches Math\'ematiques, Universit\'e de Montr\'eal, P.O. Box 6128, Centre-ville Station, Montr\'eal (Qu\'ebec), H3C 3J7, Canada}

\author{Lucia Morey}
\affiliation{Centre de Recherches Math\'ematiques, Universit\'e de Montr\'eal, P.O. Box 6128, Centre-ville Station, Montr\'eal (Qu\'ebec), H3C 3J7, Canada}

\author{Luc Vinet}
\affiliation{Centre de Recherches Math\'ematiques, Universit\'e de Montr\'eal, P.O. Box 6128, Centre-ville Station, Montr\'eal (Qu\'ebec), H3C 3J7, Canada}
\affiliation{Département de Physique, Universit\'e de Montr\'eal, P.O. Box 6128, Centre-ville Station, Montr\'eal (Qu\'ebec), H3C 3J7, Canada}

\date{\today}

\begin{abstract}
Protected zero modes are a hallmark of topological phases of matter and are exponentially localized at sharp interfaces between distinct gapped phases. We investigate how this picture changes for smooth interfaces in a broad class of inhomogeneous Su--Schrieffer--Heeger (SSH) models. Combining an exact lattice solution with an inhomogeneous Dirac description, we show that the associated Jackiw--Rebbi zero mode becomes spatially extended. For arbitrary smooth hopping profiles, its lattice extension universally scales as the square root of the system size, independently of the microscopic details of the interface. This emergent length defines a mesoscopic critical region separating two gapped phases, within which correlations decay algebraically before crossing over to exponential decay. In addition, the entanglement entropy scales as the logarithm of the emergent length near the interface, confirming the interpretation of a mesoscopic critical region. Our results establish a universal critical length governing the low-energy physics of smooth topological interfaces.
\end{abstract}
\maketitle

\paragraph{\bf Introduction.---}

Topological phases of matter evade Landau's paradigm, as they cannot be distinguished by local order parameters associated with spontaneous symmetry breaking. Instead, they are characterized by global topological invariants robust against local perturbations~\cite{hasan2010colloquium,chiu2016classification,wen2017colloquium}.
A hallmark of topological phases is the existence of gapless surface states, probed in several experimental platforms~\cite{meier2016observation,meier2018observation,angelakis2014probing,ozawa2019topological,leder2016real,cooper2019topological}. The simplest model exhibiting these phenomena is the one-dimensional Su--Schrieffer--Heeger (SSH) model~\cite{SSH}, which realizes two gapped phases distinguished by different winding numbers~\cite{asboth2016short}. 

In topological insulators such as the SSH model, interfaces between gapped phases with distinct topological invariants necessarily host gapless excitations, or zero modes. This phenomenon is described by the Jackiw--Rebbi model, in which a Dirac fermion acquires a spatially varying mass that changes sign across a domain wall~\cite{jackiw1976solitons}. For a sharp interface, the resulting zero mode is exponentially localized around the domain wall.
A qualitatively different situation arises for smooth interfaces, where the inhomogeneous mass varies slowly in space. Such interfaces have been investigated and realized in different contexts including Volkov-Pankratov states~\cite{volkov1985two,tchoumakov2017volkov,lu2019magneto,lu2020dirac,bermejo2023observation}, topological photonics~\cite{angelakis2014probing,ozawa2019topological}, and ultracold atoms~\cite{leder2016real,cooper2019topological}. Topology still guarantees the existence of the zero mode, but its density is no longer necessarily exponentially suppressed away from the interface, allowing a spatial extension much larger than the microscopic length scale. A quantitative understanding of this extension and its consequences for many-body observables remains lacking. In particular, it is unknown whether smooth topological interfaces possess a universal emergent length scale governing their low-energy physics.

Spatial inhomogeneities are ubiquitous in quantum matter and can modify the physics of otherwise translation-invariant many-body systems. This has led to approaches including conformal field theory in curved backgrounds~\cite{DSVC17,dubail2017emergence} and exactly solvable inhomogeneous lattice models~\cite{Ramirez:2015yfa,Rodriguez-Laguna:2016roi,Crampe:2019upj,Finkel:2020lgf,Finkel:2021gji,bernard2022entanglement,blanchet24Neg,bernard2025entanglementEH,bernard2025exactly}. Trapping potentials are a particularly relevant source of inhomogeneity in ultracold atoms~\cite{bloch2008many}. Remarkably, their spatial profile can generate an emergent length scale over which an otherwise critical system remains effectively critical before becoming gapped, with signatures in correlations and entanglement~\cite{campostrini2009critical,campostrini2010trap,bernard2025entanglement}. Inhomogeneous SSH chains have also been considered~\cite{mandal2024topological,rajbongshi2025topological,crampe2026Inhomogeneous,kumar2026entanglement,mandal2026quantum}, but the spatial extension of their topological zero modes and its consequences for the surrounding many-body state remain largely unexplored.

In this Letter, we investigate a broad class of inhomogeneous SSH lattice models, in which smoothly varying hopping amplitudes provide a natural realization of arbitrary smooth interfaces. Combining an exact lattice solution with an inhomogeneous Dirac description, we determine the universal spatial structure of the associated Jackiw--Rebbi zero mode. We show that, irrespective of the microscopic details, its lattice extension universally scales as $\sqrt{N}$ with the system size $N$. Moreover, this emergent length defines an effective critical region separating distinct gapped phases. Ground-state correlations decay algebraically throughout this region before crossing over to the exponential decay characteristic of gapped phases. Consistently, the entanglement entropy reveals the same effective critical region through the logarithmic scaling characteristic of one-dimensional critical free-fermion systems~\cite{vidal2003entanglement,CC04}. More generally, our results identify a universal critical length governing the low-energy physics of smooth topological interfaces.

\paragraph{\bf Model and zero mode.---} 

We consider free fermions with dimerized interactions, defined on a one-dimensional bipartite lattice where
pairs of neighbouring sites, labelled $(A,n)$ and $(B,n)$, define the unit cell~$n$. The intra- and inter-cell hopping amplitudes vary along the chain, and are denoted $v_n$ and $w_n$ respectively.
We focus on chains with open boundary conditions and odd system size $2N+1$. By convention, the last cell $n=N+1$ of the chain consists of a single site on sublattice $A$. The corresponding inhomogeneous SSH Hamiltonian reads
\begin{equation}\label{eq:inhSSHHam}
    H = \sum_{n=1}^{N} \big(v_{n}^{} c_{A,n}^\dagger c_{B,n}+w_n^{} c_{B,n}^\dagger c_{A,n+1} +\textrm{h.c.}\big), 
\end{equation}
where $c_{X,n}^{(\dagger)}$ are the fermion operators acting on the sublattice $X=A,B$ within cell $n$. In the homogeneous limit $v_n=v$ and $w_n=w$, the model reduces to the standard open SSH chain. The cases $v>w$ and $v<w$ correspond to two distinct gapped phases, separated by a critical regime for $v=w$ \cite{asboth2016short}. 

To diagonalize $H$, as for any free-fermion model, it is sufficient to diagonalize the single-particle Hamiltonian $h$, which is a $(2N+1) \times (2N+1)$ matrix encoding the coupling constants.  
The eigenproblem reads $h \Psi^{(\epsilon)} = \epsilon \Psi^{(\epsilon)}$, and we write the component of $\Psi^{(\epsilon)} $ restricted to cell $n$ as a two-component vector, $\Psi_n^{(\epsilon)} =(\psi_{A,n}^{(\epsilon)}  \ \psi_{B,n}^{(\epsilon)} )^T$. 
The model has a chiral symmetry, namely, the single-particle Hamiltonian satisfies $h\Gamma = -\Gamma h$ with $\Gamma = \textrm{diag}({1,-1,1,-1,\dots,1})$. 
Hence, we have $h\Gamma \Psi^{(\epsilon)} = -\epsilon\Gamma \Psi^{(\epsilon)}$, or $\Psi^{(-\epsilon)} = \Gamma \Psi^{(\epsilon)}$.
This implies that the spectrum of $h$ is symmetric under $\epsilon\to-\epsilon$. For odd system size, however, $h$ has odd dimension, so this symmetry necessarily enforces the existence of one vanishing eigenvalue. 
The corresponding eigenvector $\Psi^{(0)}$ is known as a \textit{zero mode}, and its components read (see Supplemental Material)
\begin{equation}\label{eq:PsinZM}
\psi_{A,n}^{(0)} = \frac{(-1)^{n-1}}{\sqrt{\mathcal{N}_{\textrm{latt}}}}  \left(\prod_{j=1}^{n-1}\frac{v_j}{w_j} \right), \qquad \psi_{B,n}^{(0)}=0,
\end{equation}
where $\mathcal{N}_{\textrm{latt}}$ is a normalisation constant. This zero mode has chirality +1, namely $\Gamma \Psi^{(0)}=+\Psi^{(0)}$, and has vanishing components on the $B$ sublattice.

For the homogeneous SSH chain, Eq.~\eqref{eq:PsinZM} implies that the zero mode is localized at the left (right) edge of the chain for $v<w$ ($v>w$), with an exponentially decaying envelope into the bulk. At the critical point $v=w$, however, the zero mode becomes fully delocalized.

\paragraph{\bf Continuum description.---}
To gain insight into the inhomogeneous case, we now turn to a continuum description of the model. We introduce the lattice spacing $a$ and the continuous coordinate $x=an$, and consider the scaling limit $a\to0$ and $N\to\infty$, with the system length $\ell=aN$ kept fixed. The lattice couplings are promoted to smooth functions, $v_n,w_n\rightarrow v(x),w(x)$, and the low-energy sector is described by the effective inhomogeneous Dirac Hamiltonian
\begin{equation}\label{eq:hDiracx}
h(x)=m(x)\sigma_x+\ir aw(x)\sigma_y\partial_x,
\end{equation}
where $m(x)=v(x)-w(x)$ and $\sigma_{x,y,z}$ are the Pauli matrices (see Supplemental Material).
The continuum Hamiltonian inherits the chiral symmetry $h(x)\sigma_z=-\sigma_zh(x)$ and admits two chiral zero modes satisfying $h(x)\Phi_\pm^{(0)}(x)=0$, and $\sigma_z\Phi_\pm^{(0)}(x)=\pm\Phi_\pm^{(0)}(x)$. Unlike the lattice model, the continuum description is insensitive to the parity of the chain and therefore admits both chiral zero modes. For the odd chains considered throughout this work, the finite lattice supports a single exact zero mode of chirality $+1$. The corresponding continuum description is
\begin{subequations}
\begin{equation}
\Phi_+^{(0)}(x)=\big(\phi_A^{(0)}(x)\ 0\big)^T,
\end{equation}
where $\phi_A^{(0)}(x)$ is the Jackiw--Rebbi mode~\cite{jackiw1976solitons}
\begin{equation}\label{eq:x0int}
\phi_A^{(0)}(x)
=\frac{1}{\sqrt{\mathcal N}}
\exp\left(\int_{\tilde x}^{x}\frac{m(s)}{aw(s)}\,\mathrm{d}s\right),
\end{equation}
\end{subequations}
and $\tilde x$ is a reference point chosen such that the exponent remains finite, and $\mathcal N$ is a normalization constant. In even chains, the lattice instead supports a pair of low-energy states with opposite chiralities and energies $\pm\delta$, where $\delta\to0$ in the thermodynamic limit. These states are described by the two continuum solutions $\Phi_\pm^{(0)}$. The continuum description and the universal scaling derived below are therefore independent of the chain parity.

\paragraph{\bf Spatial extension.---}
If $m(x)$ has a definite sign over the whole interval $x\in[0,\ell]$, the zero mode is localized at one of the two boundaries, depending on the sign of $m(x)$, as in the two gapped phases of the homogeneous SSH chain. In this case, the reference point $\tilde x$ is conveniently chosen at the boundary where the zero mode is localized, namely $\tilde x=0$ for $m(x)<0$ and $\tilde x=\ell$ for $m(x)>0$. The critical regime corresponds to $m(x)=0$, for which the zero mode is fully delocalized.

We now consider the inhomogeneous case where $m(x)$ changes sign once at $x=x_0$, with $m(x_0)=0$, $m(0)>0$ and $m(\ell)<0$. Moreover, we assume $m'(x_0)\neq 0$, where the prime denotes differentiation with respect to~$x$.
Choosing the reference point $\tilde x=x_0$ in Eq.~\eqref{eq:x0int} and linearizing the exponent around $x_0$, we obtain \cite{jackiw1976solitons,volkov1985two}
\begin{subequations}\label{eq:phiAZM}
\begin{equation}
\phi_A^{(0)}(x)=\frac{1}{(\pi \xi^2)^{1/4}}\exp\left(-\frac{(x-x_0)^2}{2a^2\xi^2}\right),
\end{equation}
where
\begin{equation}
\xi=\sqrt{\frac{w(x_0)}{a|m'(x_0)|}}.
\end{equation}
\end{subequations}
The length $a\xi$ characterizes the spatial extension of the zero mode close to the interface $x=x_0$. On a lattice, the corresponding zero mode is centred around the cell $n_0=x_0/a$, and extends over $~\xi$ lattice cells. 

We now show that the scaling of $\xi$ is universal for generic smooth inhomogeneous SSH chains admitting a continuum limit with $m'(x_0)\neq 0$. 
Let $f_{n,N}$ denote a generic lattice coupling in a chain of $N$ unit cells. The existence of a well-defined continuum limit requires the lattice profile to be invariant under a rescaling, $f_{\lambda n,\lambda N}=f_{n,N}$, for any scaling factor $\lambda>0$. Therefore, $f_{n,N}$ depends on the lattice index only through the ratio $n/N$. Passing to the continuum, the corresponding smooth function can be written as $f(x)=\tilde f\!\left(\frac{x}{\ell}\right),$ for some dimensionless function~$\tilde f$. Introducing $y=x/\ell$, it follows that $\partial_x f(x)=\ell^{-1}\partial_y \tilde f(y)$. The term $\partial_y \tilde f(y)$ does not depend on $\ell$, such that the derivative of every smooth coupling profile is of order $\ell^{-1}$.
In turn, Eq.~\eqref{eq:phiAZM} implies $\xi\sim\sqrt{\ell/a},$ or equivalently,
\begin{equation}\label{eq:xiUni}
\xi\sim\sqrt N.
\end{equation}
Hence, inhomogeneous SSH chain with a smooth and sign-changing mass profile with non-vanishing derivative at the interface hosts a spatially extended zero mode separating distinct gapped phases, whose lattice extension grows universally as~$\sqrt{N}$.
The argument immediately extends to multiple smooth interfaces, each supporting a zero mode with the same extension.

\paragraph{\bf Examples.---}

\begin{figure}[t!]
    \centering
    \includegraphics[width=1\linewidth]{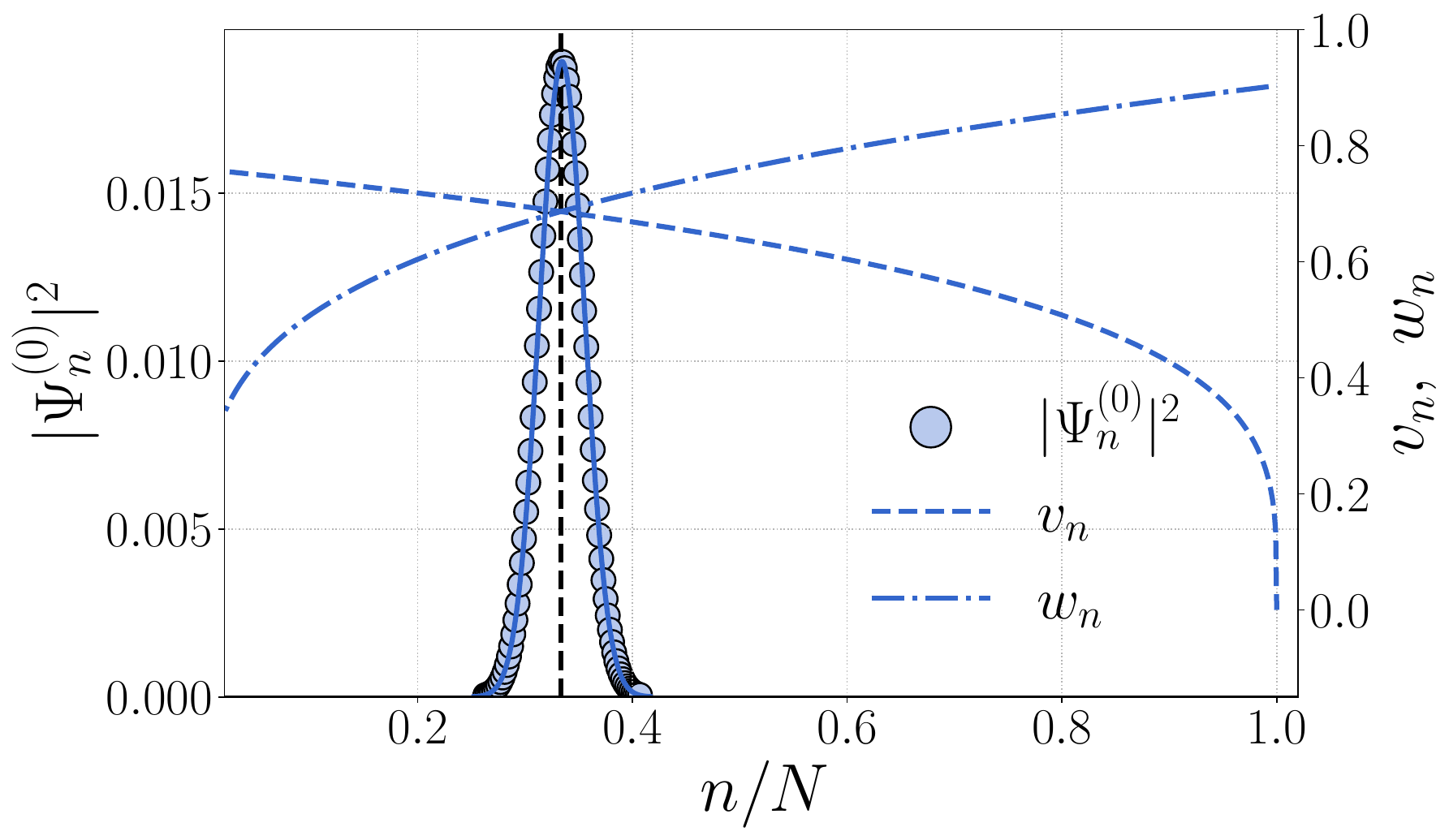}
    \caption{Spatial profile $|\Psi_n^{(0)}|^2$ of the zero mode for the power-law profile with $\alpha=1/4$ and $p=1/3$. Symbols denote the exact lattice results obtained from Eq.~\eqref{eq:PsinZM} for $N=1000$. The solid line is the continuum prediction of Eq.~\eqref{eq:phiAZM}, expressed as a function of $x/\ell=n/N$. The agreement between the lattice and continuum results is excellent. The corresponding couplings $v_n$ and $w_n$ are also shown (right axis), illustrating that the zero mode is centred at the interface $n_0/N=p=1/3$, marked by the vertical dashed line.}
    \label{fig:ZM2}
    \end{figure} 
     
To illustrate these general results, we consider two representative classes of models parametrized by $(\alpha,p)$ and $(\beta,p)$, with $0<p<1$. The first class is defined by the power-law couplings
\begin{subequations}\label{eq:vwab}
\begin{equation}
v_n^{(\alpha)}=\left(p\left(1-\frac{n}{N}\right)\right)^\alpha,\quad w_n^{(\alpha)}=\left((1-p)\frac{n}{N}\right)^\alpha,
\end{equation}
with $\alpha>0$. The second class is characterized by the Fermi--Dirac couplings
\begin{equation}
\tilde v_n^{(\beta)}=\left(\exp\!\left[\beta\left(\frac{n}{N}-p\right)\right]+1\right)^{-1},
\end{equation}
\end{subequations}
with $\tilde w_n^{(\beta)}=\tilde v_n^{(-\beta)}$ and $\beta>0$. Both families of couplings are invariant under arbitrary rescaling $(n,N)\to(\lambda n,\lambda N )$
 and therefore admit a well-defined continuum limit, obtained by identifying $n/N=x/\ell$. 

Using Eq.~\eqref{eq:phiAZM}, we find that the zero modes of both models are centred on $x_0=p\ell$, with respective spatial extensions
$\xi_\alpha=\sqrt{p(1-p)N/\alpha}$ and $\xi_\beta=\sqrt{N/\beta}$. For finite values of $\alpha$ and $\beta$, the lattice spatial extension satisfies $\xi\sim\sqrt N$ for all models, in agreement with our universal result of Eq.~\eqref{eq:xiUni}.

The limits $\alpha,\beta\to0$ recover the homogeneous critical chain, for which the zero mode is fully delocalized and $\xi\to\infty$, as expected. Conversely, the limits $\alpha,\beta\to\infty$ (at fixed $N$) produce an abrupt interface and hence fall outside the assumptions of a smooth continuum limit. For the Fermi--Dirac profile, this limit corresponds to two fully dimerized SSH chains in distinct gapped phases joined at a single bond, where the zero mode is exactly localized at the interface. These limiting cases thus provide useful consistency checks of our analysis.

In Fig.~\ref{fig:ZM2} we compare the prediction of Eq.~\eqref{eq:phiAZM} with the exact lattice result of Eq.~\eqref{eq:PsinZM} for the power-law profile with $\alpha=1/4$, $p=1/3$ and $N=1000$. The agreement between the continuum approximation and the exact lattice zero mode is excellent. We found a similarly good agreement for the Fermi--Dirac profile. Further numerical evidence is provided in the Supplemental Material, including additional examples and weakly disordered chains, as well as a verification, using the Local Topological Marker introduced in Ref.~\cite{meier2018observation}, that the interface hosting the zero mode indeed separates two gapped phases with distinct topological invariants.

\paragraph{\bf Correlations.---}

Ground-state correlation functions provide a standard probe of criticality, exhibiting algebraic decay in critical systems and exponential decay in gapped ones. 
We thus consider the ground-state correlation matrix $C^-=\sum_{\epsilon<0}\Psi^{(\epsilon)}[\Psi^{(\epsilon)}]^\dagger$. In the sublattice basis, each matrix element $C_{mn}^-$ is a $2\times2$ matrix whose entries correspond to the four possible sublattice correlations between unit cells $m$ and $n$, that we denote $C_{mn}^{-,XY}$, with $X,Y=A,B$. Using chiral symmetry together with the completeness of the single-particle eigenstates, we find 
that the correlations within each sublattice are entirely determined by the zero mode, and are thus exactly known (see Supplemental Material). 

\begin{figure}[t!]
    \centering
    \includegraphics[width=1\linewidth]{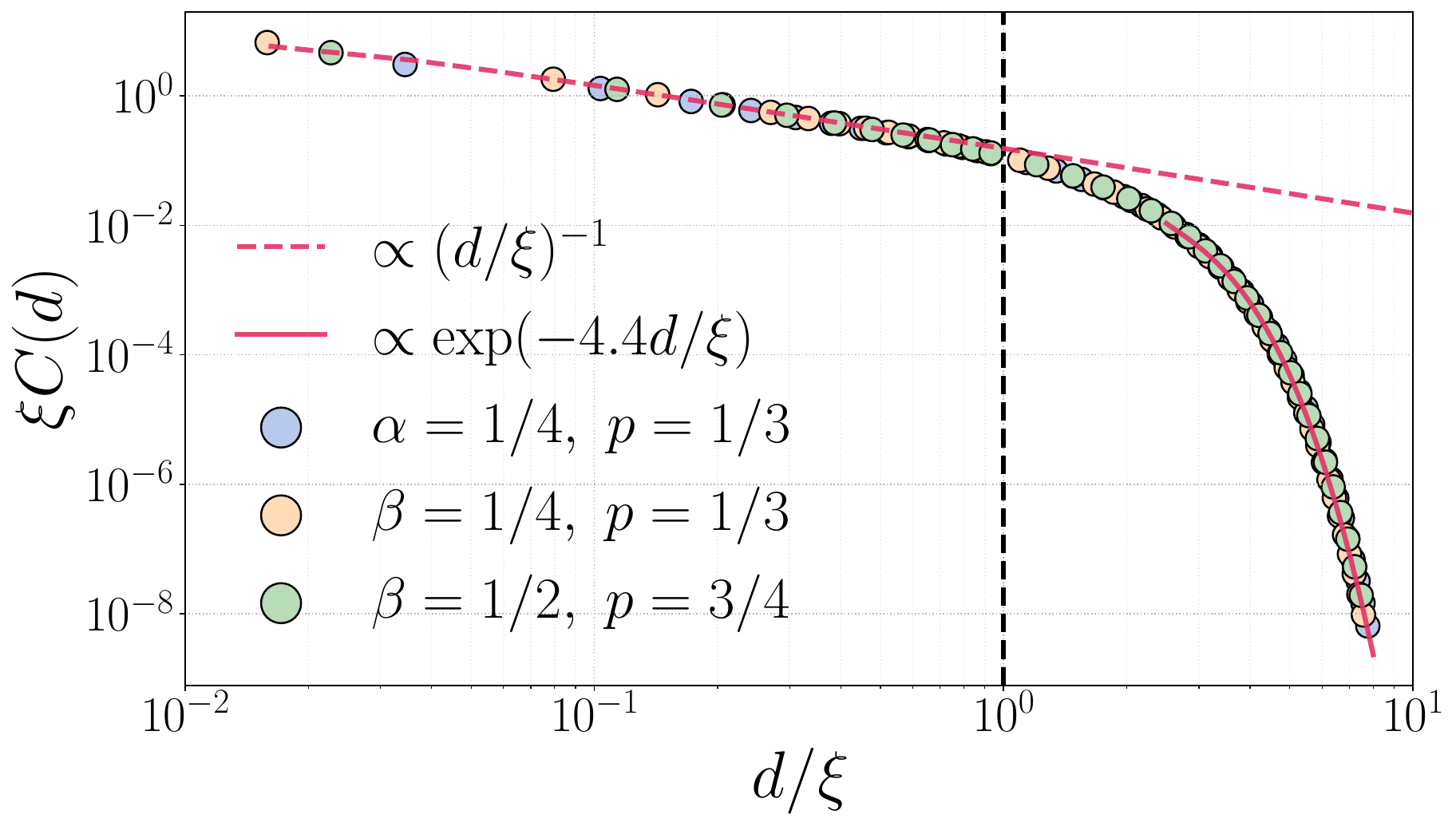}
    \caption{Rescaled correlation function $\xi C(d)$ as a function of the rescaled distance $d/\xi$ for the power-law and Fermi--Dirac profiles with $N=1000$. Symbols denote exact numerical data, and the dashed and solid lines show the algebraic and exponential fits, respectively, while the vertical dashed line indicates $d/\xi=1$.}
    \label{fig:correl}
    \end{figure}

To probe the effect of the spatial extension of the zero mode, we thus consider the $AB$ correlation $C(d)\equiv C^{-,AB}_{n_0-d/2,n_0+d/2}$ between cells symmetrically located around the interface, at an even distance $d>0$. Figure~\ref{fig:correl} shows the rescaled correlations $\xi C(d)$ as a function of the rescaled distance $d/\xi$ for the power-law and Fermi--Dirac profiles. Remarkably, the data for the two profiles collapse onto a common scaling function, demonstrating that $\xi$ sets the characteristic length scale of the correlations. For $d\ll\xi$, the correlations decay algebraically as $C(d)\sim d^{-1}$. The exponent coincides with that of the homogeneous SSH chain at criticality, providing evidence for critical free-fermion behaviour near the interface.
By contrast, for $d\gg\xi,$ the correlations become exponentially suppressed, reflecting the surrounding gapped phases. The collapse shows that this crossover occurs around $d/\xi=1$, marked by the vertical dashed line, independently of the microscopic hopping profile.

\paragraph{\bf Entanglement entropy.---}
\begin{figure}[t!]
    \centering
    \includegraphics[width=1\linewidth]{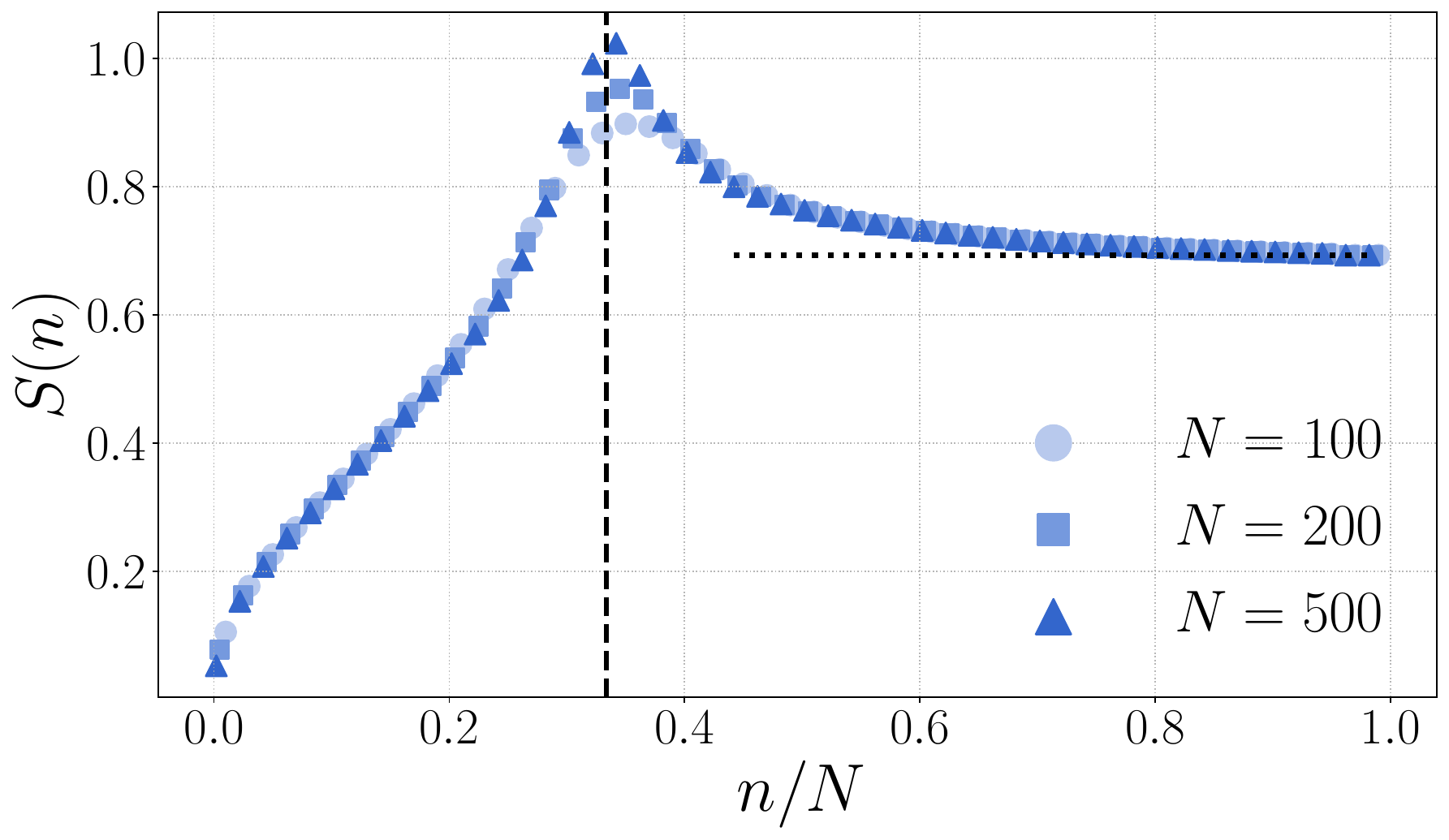}\\
    \includegraphics[width=1\linewidth]{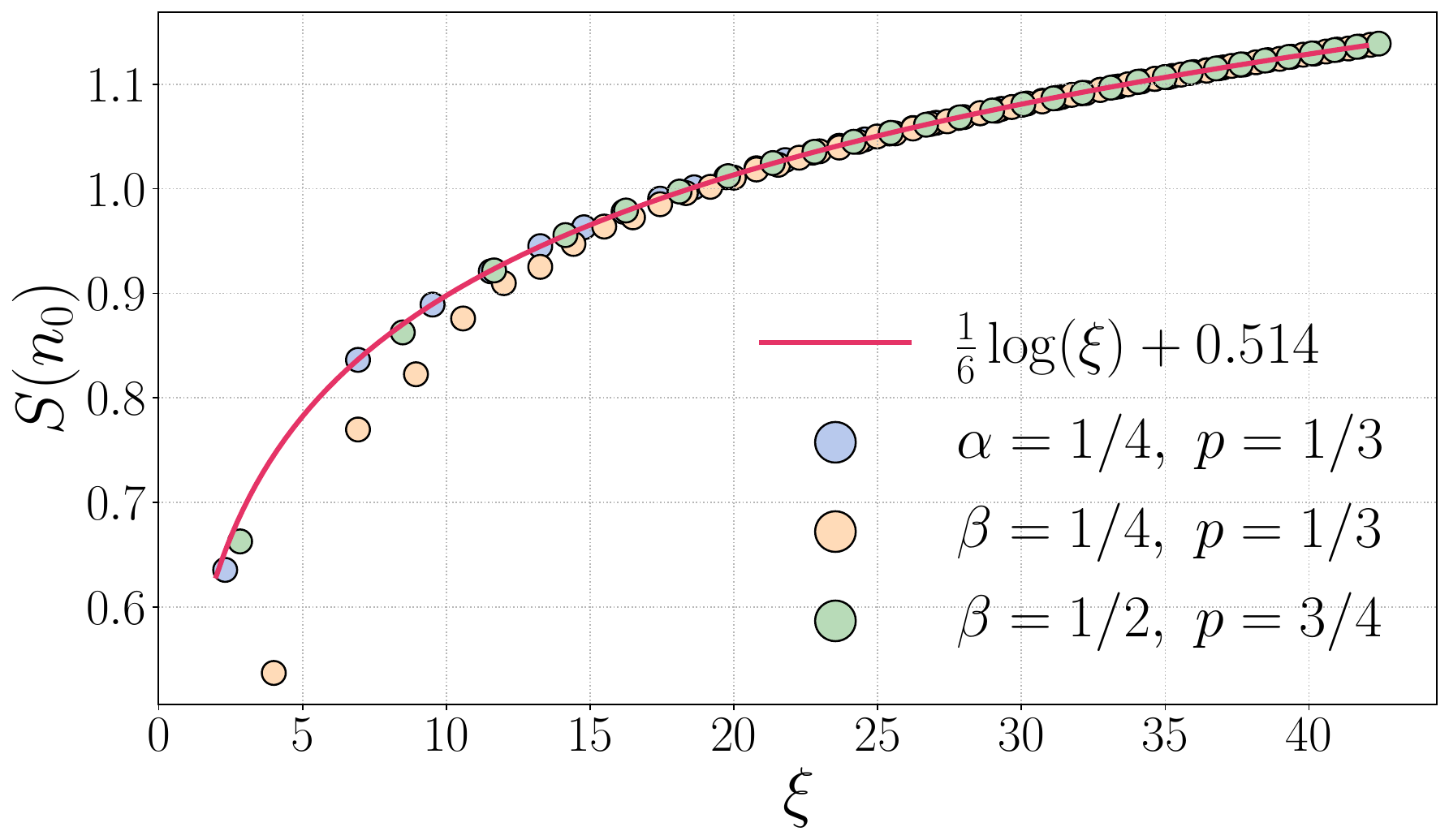}
    \caption{\textbf{Top:} Entanglement entropy $S(n)$ as a function of $n/N$ for the power-law profile with $\alpha=1/4$, $p=1/3$, and various values of $N$. Symbols are obtained by exact diagonalization of the correlation matrix. The vertical dashed line marks the interface at $n_0/N=p=1/3$, while the horizontal dotted line indicates $S=\log 2$. \textbf{Bottom:} Entanglement entropy $S(n_0)$ at the interface as a function of $\xi$ for the power-law and Fermi--Dirac profiles. Symbols are obtained by exact diagonalization, and the solid line is a numerical fit.
 }
    \label{fig:SA}
    \end{figure} 

Ground-state entanglement entropy provides another standard probe of criticality, exhibiting logarithmic scaling in critical systems and saturation in gapped ones~\cite{vidal2003entanglement,CC04}. To further characterize the physical consequences of the spatially extended zero mode, we consider
$S(n)=-\mathrm{Tr}\,\bigl(\rho_n\log\rho_n\bigr),$
where $\rho_n$ is the ground-state reduced density matrix of the first $n$ unit cells. Since the ground state is Gaussian, $S(n)$ is obtained directly from the eigenvalues of the ground-state correlation matrix restricted to these $n$ unit cells~\cite{peschel2003calculation,peschel2009reduced}. Using again chirality and the completeness of the single-particle eigenstates, we find that $S(n)$ is the same for both ground states, whether the zero mode is included or only states with strictly negative energies are occupied.
 
In the top panel of Fig.~\ref{fig:SA}, we display the entanglement entropy $S(n)$ as a function of $n/N$ for the power-law profile with $\alpha=1/4$, $p=1/3$, and various values of~$N$. Away from the interface indicated by the vertical dashed line, all curves collapse, indicating that $S(n)$ is independent of the system size, as expected in gapped phases. Furthermore, the figure clearly distinguishes the two different gapped phases. For $n>n_0$, we have $v_n<w_n$, and hence the inter-cell hopping dominates. Since the entanglement bipartition separates the unit cells $n$ and $n+1$, the cut lies on a strong link, and the entropy in this region tends to that of an EPR pair, namely $S=\log 2$, as indicated by the horizontal dotted line. By contrast, for $n<n_0$, we have $v_n>w_n$, so that the entanglement cut lies on a weak link. As a result, the entanglement entropy is small in this region and eventually vanishes toward the left end of the chain.

In the vicinity of the interface, around $n\sim n_0$, the entanglement entropy grows with $N$. To characterize this growth, we investigate the scaling of $S(n_0)=S(pN)$ with the interface width $\xi$ in the lower panel of Fig.~\ref{fig:SA}, for both the power-law and Fermi--Dirac profiles. In both cases, we obtain
\begin{equation}\label{eq:SpN}
S(n_0)=\frac{1}{6}\log \xi+0.514
\end{equation}
at leading order in $N$. For the SSH chain at the critical point, the leading term in the scaling of the entanglement entropy with the system size is $S\sim 1/6\log N$ \cite{vidal2003entanglement,CC04}. Equation~\eqref{eq:SpN} can thus be interpreted as an effective critical region of size $\xi$ centred at the interface and induced by the spatial extension of the zero mode. 

\paragraph{\bf Conclusion.---}

We have investigated topological zero modes in inhomogeneous SSH chains by combining an exact lattice analysis with an inhomogeneous Dirac description. For arbitrary smooth hopping profiles connecting two distinct gapped phases, we have shown that the interface hosts a Jackiw--Rebbi zero mode whose lattice extension universally scales as $\xi\sim\sqrt{N}$, independently of the microscopic realization of the interface.
We have further shown that this emergent length controls the low-energy physics around the interface. Ground-state correlations decay algebraically over distances shorter than $\xi$, before crossing over to the exponential decay of the surrounding gapped phases. Likewise, the entanglement entropy grows logarithmically with $\xi$ near the interface, consistent with an effective critical region, while away from the interface it becomes independent of $N$, as expected in gapped phases.

This mechanism is reminiscent of the effective critical regions generated by trapping potentials in free-fermion and free-boson systems~\cite{campostrini2009critical,campostrini2010trap,bernard2025entanglement}, but arises here from a fundamentally different origin: the spatial extension of a topologically protected zero mode. This provides a new mechanism for generating an effective critical region at a topological interface, linking the spatial structure of a protected boundary state to the many-body correlations and entanglement of the surrounding system. Since the scaling of $\xi$ is independent of the microscopic hopping profile, we expect these results to extend beyond the specific models considered here. 

Smooth topological interfaces have been proposed in photonic platforms~\cite{angelakis2014probing} and realized experimentally in ultracold atoms~\cite{leder2016real}, while recent photonic experiments have demonstrated controlled Dirac mass engineering and topological boundary states~\cite{yu2024dirac}. These platforms thus provide promising settings in which to probe the spatial extension of topological zero modes and its consequences for correlations and entanglement.
More broadly, it would be interesting to investigate whether similar physics arises in other topological lattice models, and whether the spatial extension of their zero modes leads to analogous effects in nonequilibrium dynamics.
    
\paragraph{\bf Acknowledgments.---} L.~Vinet is funded in part by a Discovery Grant from the Natural Sciences and Engineering Research Council (NSERC) of Canada. Q. Labriet and L. Morey enjoy postdoctoral fellowships provided by this grant. G.~Parez thanks Clément Berthiere for useful discussions.

\providecommand{\href}[2]{#2}\begingroup\raggedright\endgroup

%%%%%%%%%%%%%%%%%%%%%%%%%%%%%%%%%%%%%%%%%%%%%%%%%%%%%%%%%%%%%%%%%
%%%%%%%%%%%%%%%%%%%%%%%%%%%%%%%%%%%%%%%%%%%%%%%%%%%%%%%%%%%%%%%%%

\onecolumngrid
\setcounter{secnumdepth}{3}
\clearpage
\begin{center}

\textbf{\large Supplemental Material: Universal scaling of spatially extended zero modes in inhomogeneous SSH chains}\\[0.4cm]

\end{center}

\vspace{0.5cm}

\setcounter{section}{0}
\setcounter{equation}{0}
\setcounter{figure}{0}
\setcounter{page}{1}
\thispagestyle{empty}
\setcounter{table}{0}
\makeatletter
\renewcommand{\thesection}{S\arabic{section}}
\renewcommand{\theequation}{S\arabic{equation}}
\renewcommand{\thefigure}{S\arabic{figure}}
\renewcommand{\bibnumfmt}[1]{[S#1]}

\begin{center}
\textbf{\large Contents}
\end{center}

\renewcommand{\arraystretch}{1.5}

\begin{tabular*}{\textwidth}{@{\extracolsep{\fill}}lr}
\hyperref[sec:zm]{S1 \hspace{0.4em} Single-particle Hamiltonian and zero mode} & 1\\
\hyperref[sec:continuum]{S2 \hspace{0.4em} Continuum description and effective Dirac equation} & 2\\
\hyperref[sec:corr]{S3 \hspace{0.4em} Correlation matrices} & 3\\
\hyperref[sec:ltm]{S4 \hspace{0.4em} Local topological marker} & 4\\
\hyperref[sec:robust]{S5 \hspace{0.4em} Robustness of the results} & 5\\
\qquad\hyperref[subsec:profiles]{A \hspace{0.4em} Additional tests for power-law and Fermi--Dirac profiles} & 5\\
\qquad\hyperref[subsec:disorder]{B \hspace{0.4em} Weak disorder} & 5\\
\end{tabular*}

\section{Single-particle Hamiltonian and zero mode }\label{sec:zm}

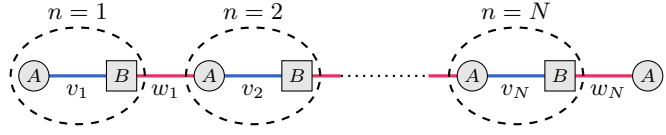
\begin{figure}[t!]
\begin{center}
\begin{tikzpicture}[scale=1.16]

\draw[-,blueG, very thick] (0,0)--(1,0);
\draw[-,redG, very thick] (1,0)--(2,0);
\draw[-,blueG,very thick] (2,0)--(3,0);
\draw[-,redG,very thick] (3,0)--(3.5,0);
\draw[-,redG,very thick] (4.5,0)--(5,0);
\draw[-,blueG,very thick] (5,0)--(6,0);
\draw[-,redG,very thick] (6,0)--(7,0);

\draw[dotted, thick] (3.5,0)--(4.5,0);

\foreach \s in {0,2,5,7} {
  \node[
    circle,
    draw,
    fill=gray!20,
    minimum size=4mm,
    inner sep=0pt
  ] at (\s,0) {\scriptsize $A$};
}

\foreach \s in {1,3,6} {
  \node[
    rectangle,
    draw,
    fill=gray!20,
    minimum size=4mm,
    inner sep=0pt
  ] at (\s,0) {\scriptsize $B$};
}

\draw[thick, dashed] (0.5,0) ellipse (0.765 and 0.55);
\draw[thick, dashed] (2.5,0) ellipse (0.765 and 0.55);
\draw[thick, dashed] (5.5,0) ellipse (0.765 and 0.55);

\node[above] at (0.5,0.55) {$n = 1$};
\node[above] at (2.5,0.55) {$n = 2$};
\node[above] at (5.5,0.55) {$n = N$};

\node[below] at (0.5,0) {$v_1$};
\node[below] at (1.5,0) {$w_1$};
\node[below] at (2.5,0) {$v_2$};
\node[below] at (5.5,0) {$v_N$};
\node[below] at (6.55,0) {$w_N$};

\end{tikzpicture}

\end{center}
\caption{
Representation of an inhomogeneous SSH model of size $2N+1$ with varying intra- and inter-cell hopping amplitudes $v_n$ and $w_n$. Sites in sublattice $A$ ($B$) are represented by a circle (square). }
\label{fig:sshInh}  
\end{figure}

The single-particle Hamiltonian $h$ associated to the full free-fermion Hamiltonian given in the main text is the following $(2N+1)\times(2N+1)$ tridiagonal matrix,
\begin{equation}\label{eq:h}
h =
\begin{pmatrix}
0 & v_1 \\
v_1 & 0 & w_1 \\
& w_1 & 0 & v_2 \\
& & v_2 & 0 & \ddots \\
& & & \ddots & \ddots & w_N \\
& & & & w_N & 0
\end{pmatrix}.
\end{equation}
We illustrate this model in Fig.~\ref{fig:sshInh}

The eigenvector of $h$ associated to the energy $\epsilon$ is denoted $\Psi^{(\epsilon)}$, and we recall that we write the component of $\Psi^{(\epsilon)} $ restricted to cell $n$ as a two-component vector, $\Psi_n^{(\epsilon)} =(\psi_{A,n}^{(\epsilon)}  \ \psi_{B,n}^{(\epsilon)} )^T$ for $n=1,2,\dots, N$. Moreover, the last cell $n=N+1$ consists of a single site on sublattice $A$, namely $\Psi_{N+1}^{(\epsilon)} =\psi_{A,N+1}^{(\epsilon)}$.

To obtain the exact solution for the components of the zero mode, we solve the eigenproblem $h\Psi^{(0)} = 0$. In coordinates, this translates to the following recurrence relation
\begin{equation}
\begin{cases}
v_{n}\psi_{A,n}^{(0)}+w_{n}\psi_{A,n+1}^{(0)} =0, \quad &n=1,2,\dots, N, \\[0.15cm]
w_{n}\psi_{B,n}^{(0)}+v_{n+1}\psi_{B,n+1}^{(0)} = 0,  \quad &n=1,2,\dots, N-1,
\end{cases}
\end{equation}
with the boundary conditions $v_1\psi_{B,1}^{(0)}=0$ and $w_N \psi_{B,N}^{(0)}=0$. The solution of this recurrence relation is $\psi_{B,n}^{(0)}=0$ for $n=1,2,\dots, N,$ and 
\begin{equation}\label{eq:psiA0}
\psi_{A,n}^{(0)} = (-1)^{n-1} \left(\prod_{j=1}^{n-1}\frac{v_j}{w_j} \right)\psi_{A,1}^{(0)}, \quad n=2,3,\dots,N+1. 
\end{equation}
The value of $\psi_{A,1}^{(0)}$ is arbitrary and we choose it such that the state is normalised. Hence, we recast Eq.~\eqref{eq:psiA0} as
\begin{equation}\label{eq:PsinZM_SM}
\psi_{A,n}^{(0)} = \frac{(-1)^{n-1}}{\sqrt{\mathcal{N}_{\textrm{latt}}}}  \left(\prod_{j=1}^{n-1}\frac{v_j}{w_j} \right)
\end{equation}
in the main text, where the normalisation constant $\mathcal{N}_{\textrm{latt}}$ ensures that $|\Psi^{(0)}|^2=1$. 

\section{Continuum description and effective Dirac Hamiltonian}\label{sec:continuum}

In this section, we derive the continuum inhomogeneous Hamiltonian given in Eq~\eqref{eq:hDiracx}. We introduce the lattice spacing $a$ and the coordinate $x=na$. We then consider the scaling limit where $a\to 0$ and $N \to \infty$ such that the total length of the chain $\ell=aN$ remains constant. In this limit, the variable $x\in[0,\ell]$ becomes a continuous variable and the idea is then to promote lattice couplings to continuous functions $f_n \to f(x)$ with $f=v,w$. However, for this operation to be meaningful and lead to a well-defined continuous model, the constants need to vary smoothly in space. At the level of the lattice, we impose 
\begin{equation}\label{eq:fnp1fn}
\left|\frac{f_{n+1}-f_n}{f_n}\right|\ll 1. 
\end{equation}
In the examples considered in the main text, this condition does not always hold at the edges of the chain. However, this does not affect the results in the bulk. In continuous variables, we have $f_{n+1}\to f(x+a) = f(x)+a\partial_x f(x)+\mathcal{O}(a^2)$. Neglecting the terms of order $a^2$, the lattice condition of Eq.~\eqref{eq:fnp1fn} gives
\begin{equation}\label{eq:smoothVW}
\left|\frac{a\partial_xf(x)}{f(x)}\right| \ll1,
\end{equation}
for $x\in[0,\ell]$. This is the criteria that defines smoothly-varying functions which we consider in this work.  

To proceed, we also need a continuous description for the components of the eigenvectors of $h$. However, one cannot directly identify $\psi_{A,n}^{(\epsilon)}$ to a function $\psi_{A}^{(\epsilon)}(x)$. Indeed, looking at the zero mode components in Eq.~\eqref{eq:PsinZM_SM}, while the ratios $v_j/w_j$ yield smooth functions in the scaling limit, the factor $(-1)^n$ does not. Instead, the lattice zero mode consists of a rapidly oscillating factor multiplied by a slowly varying envelope. This feature is shared by all low-energy eigenstates. We therefore introduce
\begin{equation}
\Phi_n^{(\epsilon)}\equiv (-1)^n\Psi_n^{(\epsilon)}=
\begin{pmatrix}
\phi_{A,n}^{(\epsilon)}\\
\phi_{B,n}^{(\epsilon)}
\end{pmatrix},
\end{equation}
where $\phi_{A,n}^{(\epsilon)}$ and $\phi_{B,n}^{(\epsilon)}$ vary slowly from one unit cell to the next and can thus be promoted to continuous functions,
\begin{equation}
\Phi_n^{(\epsilon)}\longrightarrow \Phi^{(\epsilon)}(x)=
\begin{pmatrix}
\phi_A^{(\epsilon)}(x)\\
\phi_B^{(\epsilon)}(x)
\end{pmatrix}.
\end{equation}

The bulk recurrence relation for $\Phi_n^{(\epsilon)}$ reads
\begin{equation}
\begin{cases}
-w_{n-1}\phi_{B,n-1}^{(\epsilon)}+v_{n}\phi_{B,n}^{(\epsilon)} = \epsilon \phi_{A,n}^{(\epsilon)}, \\[0.15cm]
v_{n}\phi_{A,n}^{(\epsilon)}-w_{n}\phi_{A,n+1}^{(\epsilon)} =\epsilon \phi_{B,n}^{(\epsilon)}.
\end{cases}
\end{equation}
We identify lattice quantities with their continuous counterparts and expand them in $a$. Neglecting the terms of order~$a^2$, we find
\begin{equation}
\begin{cases}
-[w(x)-a\partial_xw(x)][\phi_{B}^{(\epsilon)}(x)-a\partial_x\phi_B^{(\epsilon)}(x)]+v(x)\phi_{B}^{(\epsilon)}(x) = \epsilon \phi_{A}^{(\epsilon)}(x), \\[0.15cm]
v(x)\phi_{A}^{(\epsilon)}(x)-w(x)[\phi_{A}^{(\epsilon)}(x)+a\partial_x\phi_A^{(\epsilon)}(x) ]=\epsilon \phi_{B}^{(\epsilon)}(x).
\end{cases}
\end{equation}
To proceed, we use the regularity condition of Eq.~\eqref{eq:smoothVW} and neglect $a\partial_xw(x)$ compared to $w(x)$. This gives
 \begin{equation}
\begin{cases}
m(x)\phi_{B}^{(\epsilon)}(x)+a w(x)\partial_x\phi_B^{(\epsilon)}(x) = \epsilon \phi_{A}^{(\epsilon)}(x), \\[0.15cm]
m(x)\phi_{A}^{(\epsilon)}(x)-aw(x)\partial_x\phi_A^{(\epsilon)}(x) =\epsilon \phi_{B}^{(\epsilon)}(x),
\end{cases}
\end{equation}
where we introduce $m(x)=v(x)-w(x)$. 
This relation can be recast as $h(x)\Phi^{(\epsilon)}(x)=\epsilon\Phi^{(\epsilon)}(x)$, where $h(x)$ is the inhomogeneous Dirac Hamiltonian~\eqref{eq:hDiracx}.

\section{Correlation matrices}\label{sec:corr}

In the free-fermion many-body ground state, the contribution of each occupied single-particle eigenstate of $h$ to the correlation matrix is the projector $\Psi^{(\epsilon)}[\Psi^{(\epsilon)}]^\dagger$. 
We therefore introduce the projectors onto the positive- and negative-energy eigenspaces,
\begin{equation}\label{eq:Cpm}
C^+=\sum_{\epsilon>0}\Psi^{(\epsilon)}[\Psi^{(\epsilon)}]^\dagger, \qquad C^-=\sum_{\epsilon<0}\Psi^{(\epsilon)}[\Psi^{(\epsilon)}]^\dagger,
\end{equation}
together with the projector onto the zero mode,
\begin{equation}
C^0=\Psi^{(0)}[\Psi^{(0)}]^\dagger.
\end{equation}
The presence of the zero mode gives rise to a twofold-degenerate many-body ground state, whose correlation matrices are therefore $C^-$, and $
C^-+C^0.$

In the sublattice basis, each matrix element is itself a $2\times2$ block,
\begin{equation}
C^\gamma_{mn}
=
\begin{pmatrix}
C^{\gamma,AA}_{mn} &
C^{\gamma,AB}_{mn}
\\[0.15cm]
C^{\gamma,BA}_{mn} &
C^{\gamma,BB}_{mn}
\end{pmatrix},
\qquad
\gamma=0,\pm.
\end{equation}
Using the chiral symmetry relation
\begin{equation}
\Psi^{(-\epsilon)}=\Gamma\Psi^{(\epsilon)},
\end{equation}
one immediately finds
\begin{equation}
C^-=\Gamma C^+\Gamma
\end{equation}
where we recall that $\Gamma = \textrm{diag}({1,-1,1,-1,\dots,1})$. 
This implies
\begin{equation}
C^{-,AA}_{mn}=C^{+,AA}_{mn},
\quad 
C^{-,BB}_{mn}=C^{+,BB}_{mn},
\quad 
C^{-,AB}_{mn}=-C^{+,AB}_{mn},
\quad 
C^{-,BA}_{mn}=-C^{+,BA}_{mn}.
\end{equation}

Since the zero mode resides entirely on sublattice $A$, we have
\begin{equation}
C^{0,AB}_{mn}=C^{0,BA}_{mn}=C^{0,BB}_{mn}=0,
\end{equation}
while
\begin{equation}
C^{0,AA}_{mn}=\psi^{(0)}_{A,m}\psi^{(0)}_{A,n}.
\end{equation}
Using Eq.~\eqref{eq:PsinZM}, this becomes
\begin{equation}\label{eq:Czm}
C_{mn}^{0,AA}=\frac{(-1)^{m+n}}{\mathcal{N}_{\textrm{latt}}}\left(\prod_{j=1}^{m-1}\frac{v_j}{w_j}\right)\left(\prod_{k=1}^{n-1}\frac{v_k}{w_k}
\right),
\end{equation}
whose continuum approximation follows directly from Eq.~\eqref{eq:phiAZM} as the product of two Gaussian envelopes.

Completeness of the single-particle eigenstates implies
\begin{equation}
C^-+C^++C^0=\mathbb I.
\end{equation}
Together with the relations derived above, this identity yields
\begin{equation}
C^{-,AA}_{mn}=\frac12\left(\delta_{mn}-C^{0,AA}_{mn}\right),
\quad 
C^{-,BB}_{mn}=\frac12\delta_{mn}.
\end{equation}
Hence, the correlations between the same sublattice are completely fixed by the zero mode and are known analytically with Eq.~\eqref{eq:Czm}. The only non-trivial ground-state correlations are therefore the correlators $C^{-,AB}_{mn}$, which we investigate in the main text. 

\section{Local topological marker}\label{sec:ltm}

In this section, we verify that the interface hosting the zero mode is indeed separating two gapped phases with distinct topological invariants. For translation-invariant systems, such as the homogeneous SSH chain with periodic boundary conditions, a standard topological marker is the winding number. It is an integral over momentum space which can only take integer values, differentiating distinct gapped phases. Here, we consider models which are not translation-invariant, and hence the winding number is not a convenient marker. Instead, we use the Local Topological Marker (LTM) introduced in~\cite{meier2018observation}. 

The LTM is constructed in part from the projectors onto the positive and negative energy eigenspaces of the single-particle Hamiltonian, which are nothing but the correlation matrices $C^+$ and $C^-$ introduced in Eq.~\eqref{eq:Cpm}, respectively. We further introduce the projectors onto the two sublattices,
\begin{equation}
\Gamma_A=\mathrm{diag}(1,0,1,0,\ldots,1),\qquad
\Gamma_B=\mathrm{diag}(0,1,0,1,\ldots,0),
\end{equation}
together with the position operator
\begin{equation}
X=\mathrm{diag}(1,2,\ldots,2N+1).
\end{equation}
Following Ref.~\cite{meier2018observation}, we define
\begin{equation}
Q=C^+-C^-,
\end{equation}
and introduce its off-diagonal blocks
\begin{equation}
Q_{AB}=\Gamma_AQ\Gamma_B,\qquad
Q_{BA}=\Gamma_BQ\Gamma_A.
\end{equation}
We finally introduce the matrix
\begin{equation}
\nu^{\textrm{mat}}=\frac12\left(Q_{BA}[X,Q_{AB}]+Q_{AB}[Q_{BA},X]\right).
\end{equation}
The LTM associated with unit cell $n$, which we denote by $\nu_n$, is obtained by summing the diagonal entries of $\nu^{\textrm{mat}}$ over the two sites belonging to that cell,
\begin{equation}
\nu_n=\nu^{\textrm{mat}}_{2n-1,2n-1}+\nu^{\textrm{mat}}_{2n,2n}.
\end{equation}

\begin{figure}[t!]
    \centering
    \includegraphics[width=0.55\linewidth]{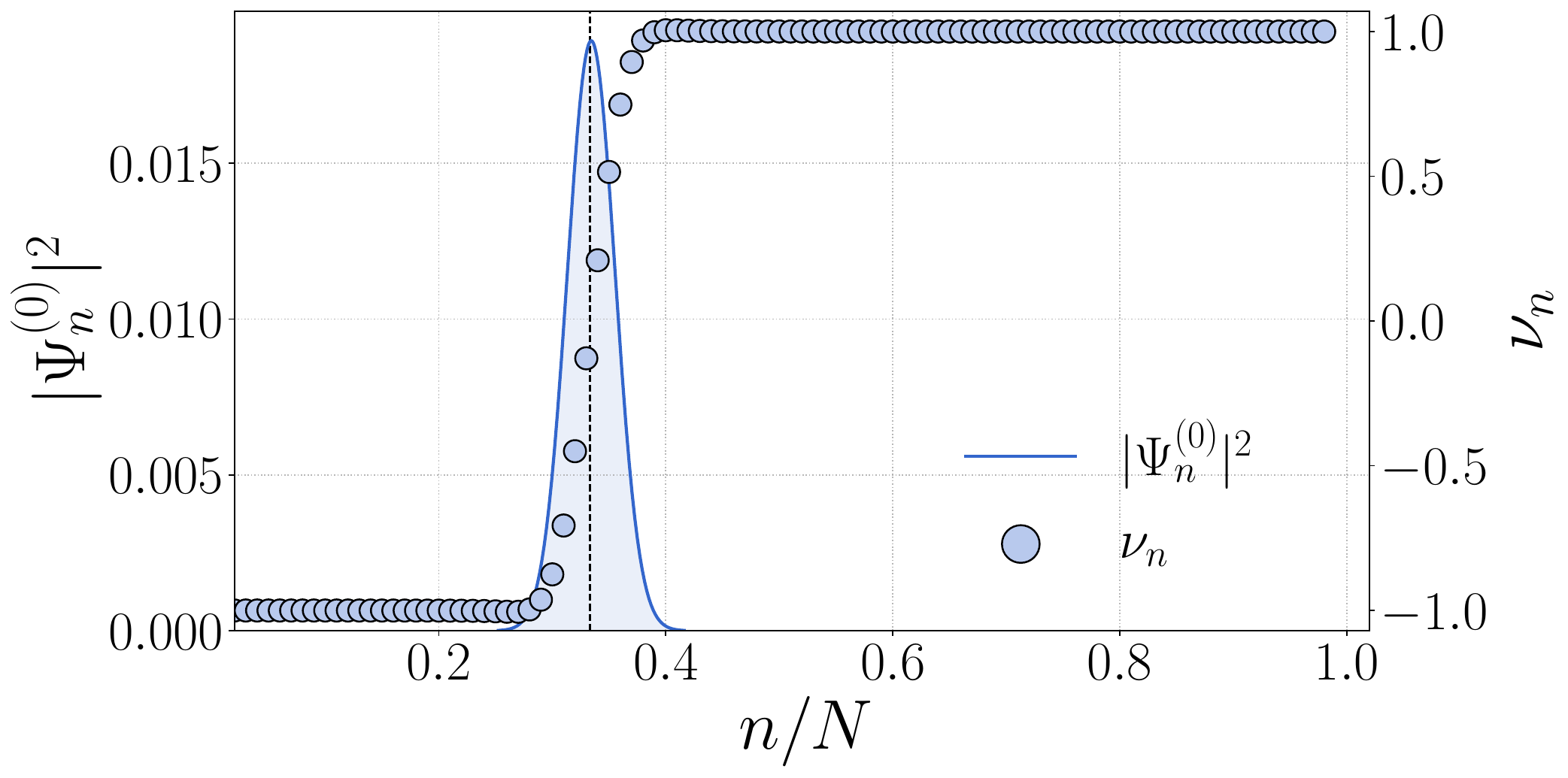}
    \caption{
Local topological marker $\nu_n$ (symbols) for the power-law profile with $\alpha=1/4$, $p=1/3$ and $N=1000$, together with the continuum prediction for the zero-mode density $|\Psi_n^{(0)}|^2$ (solid line). The vertical dashed line indicates the interface located at $n_0/N=p=1/3$. The LTM interpolates between the two quantized topological phases across the same mesoscopic region where the zero mode is spatially extended.
}
    \label{fig:nun_ZM2_SM}
    \end{figure}

We now evaluate the LTM for the power-law profile with $\alpha=1/4$ and $p=1/3$, corresponding to the example discussed in the main text. The result is shown in Fig.~\ref{fig:nun_ZM2_SM}, together with the continuum prediction for the zero-mode profile. The LTM interpolates between the quantized values $\nu_n=-1$ and $\nu_n=1$, confirming that the interface hosting the zero mode indeed separates two distinct gapped phases. Moreover, the crossover of the LTM occurs precisely over the spatial region where the zero-mode density is extended. 

\section{Robustness of the results}\label{sec:robust}

\subsection{Additional tests for power-law and Fermi-Dirac profiles}\label{subsec:profiles}

\begin{figure}[t!]
    \centering
    \includegraphics[width=0.48\linewidth]{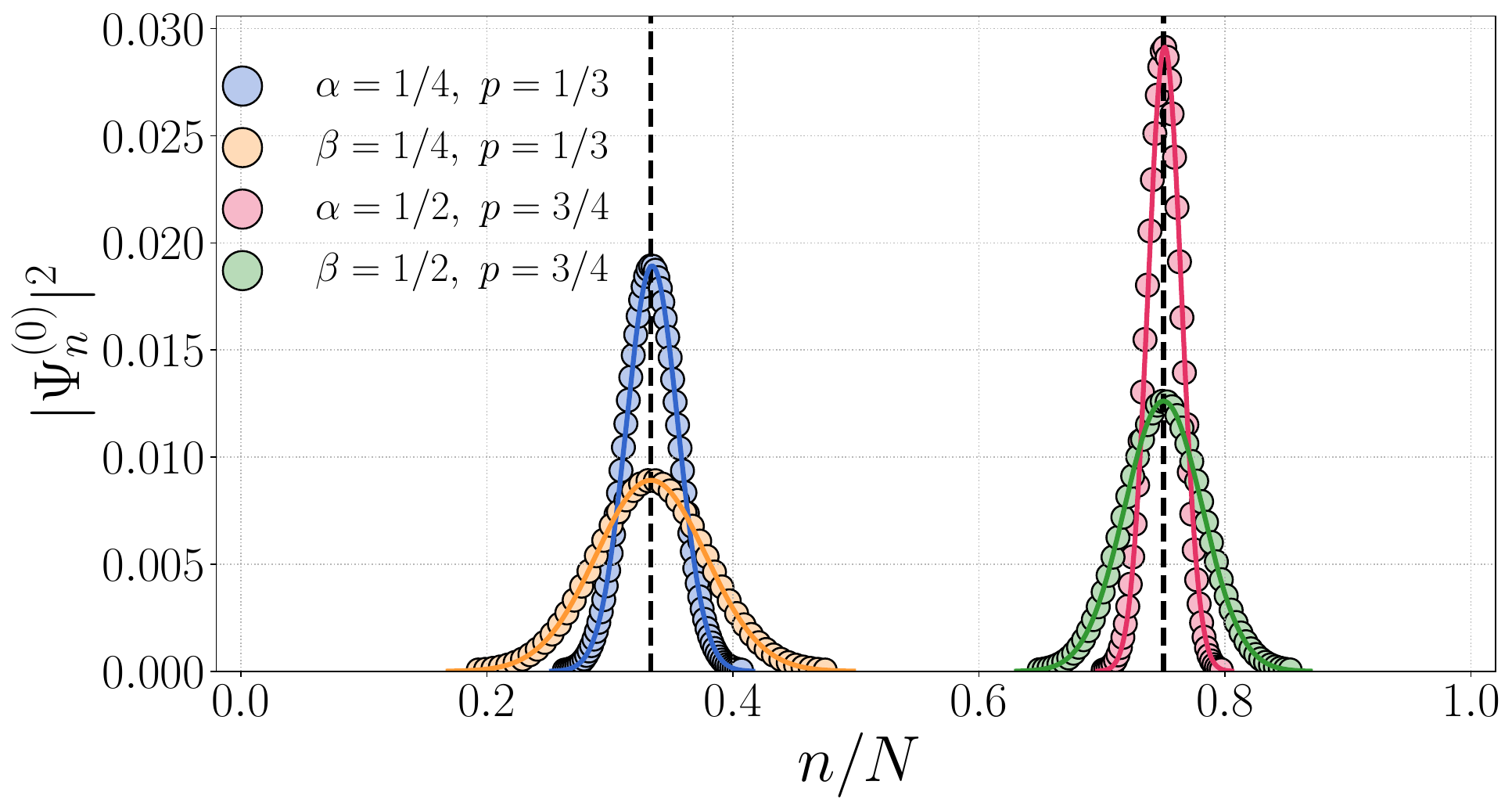}
    \includegraphics[width=0.48\linewidth]{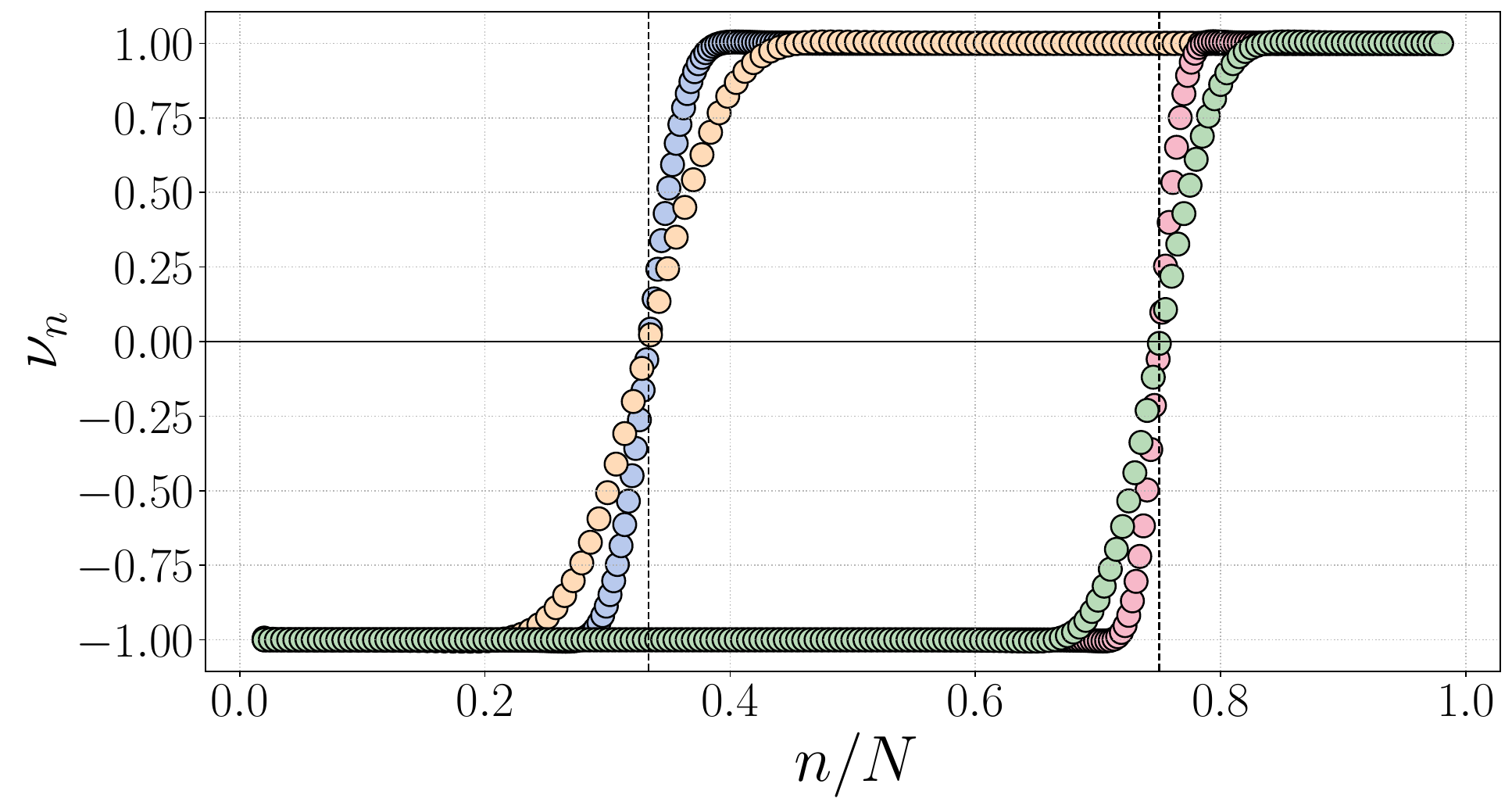}

    \caption{
   \textbf{Left:} Spatial profile $|\Psi_n^{(0)}|^2$ of the zero mode for power-law and Fermi--Dirac profiles with various values of $p$, $\alpha$ and $\beta$. The symbols are the exact lattice results obtained from Eq.~\eqref{eq:PsinZM} for $N=1000$, while the solid lines are the continuum prediction of Eq.~\eqref{eq:phiAZM} expressed as a function of $x/\ell=n/N$. The vertical dashed lines indicate the interface positions $n_0/N=p$. The agreement between the continuum approximation and the exact lattice results is excellent for all parameter choices. \textbf{Right:}  Local topological marker $\nu_n$ for the same models as in the left panel. In every case, the LTM interpolates smoothly between the quantized values $-1$ and $1$, confirming that the zero mode is located at the interface between two distinct gapped phases. The width of the crossover closely follows the spatial extension of the corresponding zero mode.}
    \label{fig:ZM2_nun_many_SM}
\end{figure}

We first consider additional power-law and Fermi--Dirac profiles, in order to assess the robustness of the continuum description and the associated topological properties. The left panel of Fig.~\ref{fig:ZM2_nun_many_SM} compares the continuum prediction for the zero-mode profile with the exact lattice result, while the right panel shows the corresponding LTM. In all cases, the agreement between the continuum and lattice results remains excellent. Moreover, the LTM always interpolates between the quantized values $-1$ and $1$ across the same mesoscopic region where the corresponding zero mode is spatially extended, confirming that it indeed separates two distinct gapped phases.

\subsection{Weak disorder}\label{subsec:disorder}

\begin{figure}[t!]
    \centering
    \includegraphics[width=0.55\linewidth]{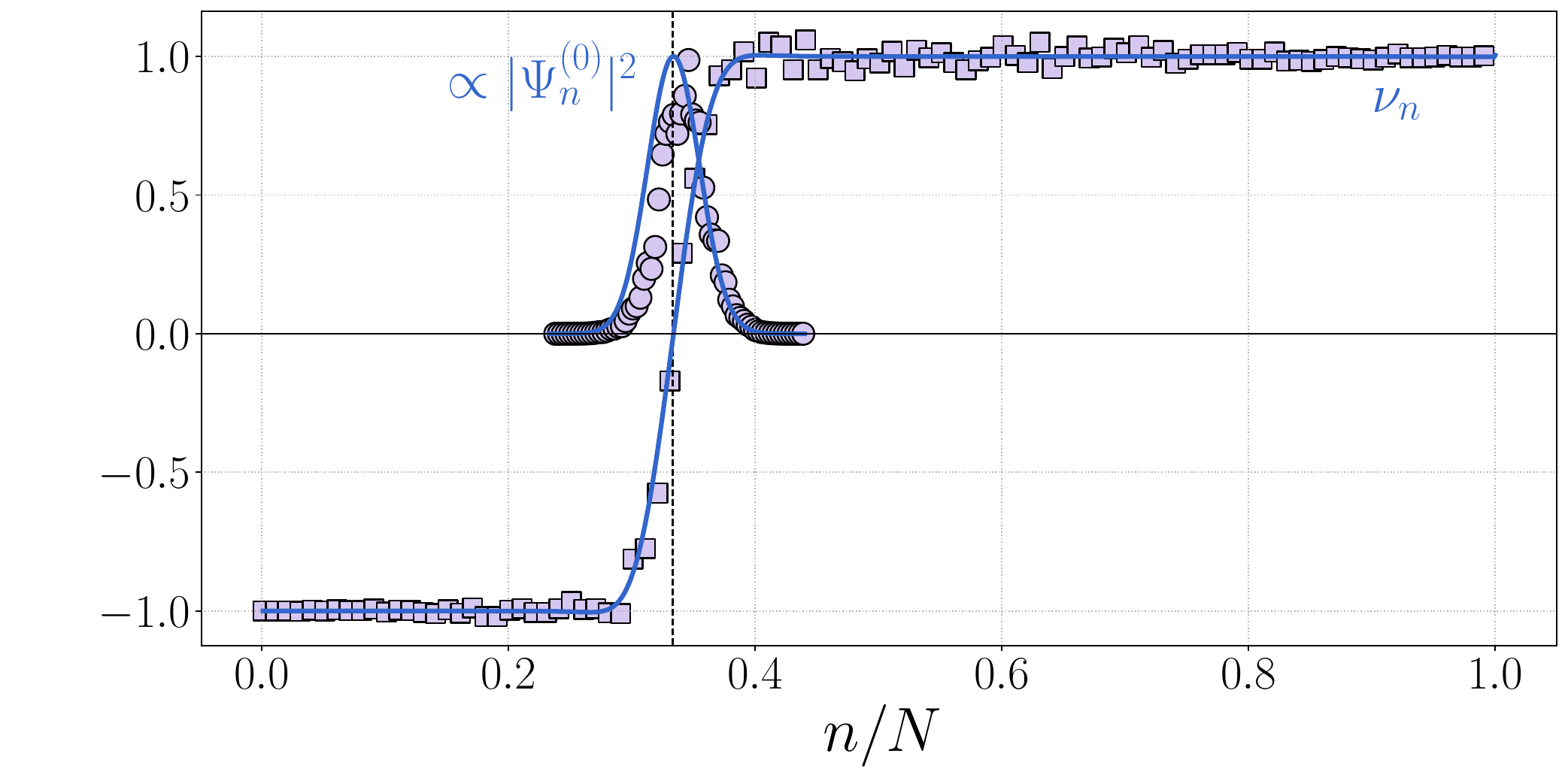}
    \caption{
    Spatial profile $|\Psi_n^{(0)}|^2$ of the zero mode (circles) and local topological marker $\nu_n$ (squares) for a single disorder realization with $W=0.05$, together with the corresponding clean results (solid lines), for the power-law profile with $\alpha=1/4$, $p=1/3$, and $N=1000$. The zero-mode density has been rescaled to share the same vertical axis as the local topological marker. Weak disorder leaves both the topological interface and the spatial extension of the zero mode essentially unchanged.
    }
    \label{fig:ZM2_nun_dis_SM}
\end{figure}

To further assess the robustness of our results, we consider inhomogeneous SSH chains with weak disorder. If $v_n$ and $w_n$ denote the couplings of the clean model, we introduce disorder through
\begin{equation}
v_n^{\mathrm{dis}} = v_n(1+W\eta_n),\qquad
w_n^{\mathrm{dis}} = w_n(1+W\tilde{\eta}_n),
\end{equation}
where $\eta_n$ and $\tilde{\eta}_n$ are independent random variables uniformly distributed in the interval $[-1,1]$, and $W$ controls the disorder strength. Since chiral symmetry is preserved, the disordered Hamiltonian still admits an exact zero mode, whose wavefunction is still given by Eq.~\eqref{eq:PsinZM} after replacing the clean couplings by their disordered counterparts.

As a first illustration, Fig.~\ref{fig:ZM2_nun_dis_SM} shows the zero-mode density together with the local topological marker for a single disorder realization with $W=0.05$, using the power-law profile with $\alpha=1/4$, $p=1/3$, and $N=1000$. The zero-mode density has been rescaled to share the same vertical axis as the LTM. The symbols correspond to the disordered chain, whereas the solid lines show the clean case. Despite weak disorder, the interface between the two topological phases remains clearly visible and the zero mode stays localized around it with essentially the same spatial extension as in the clean system.

\begin{figure}[t!]
    \centering
    \includegraphics[width=0.48\linewidth]{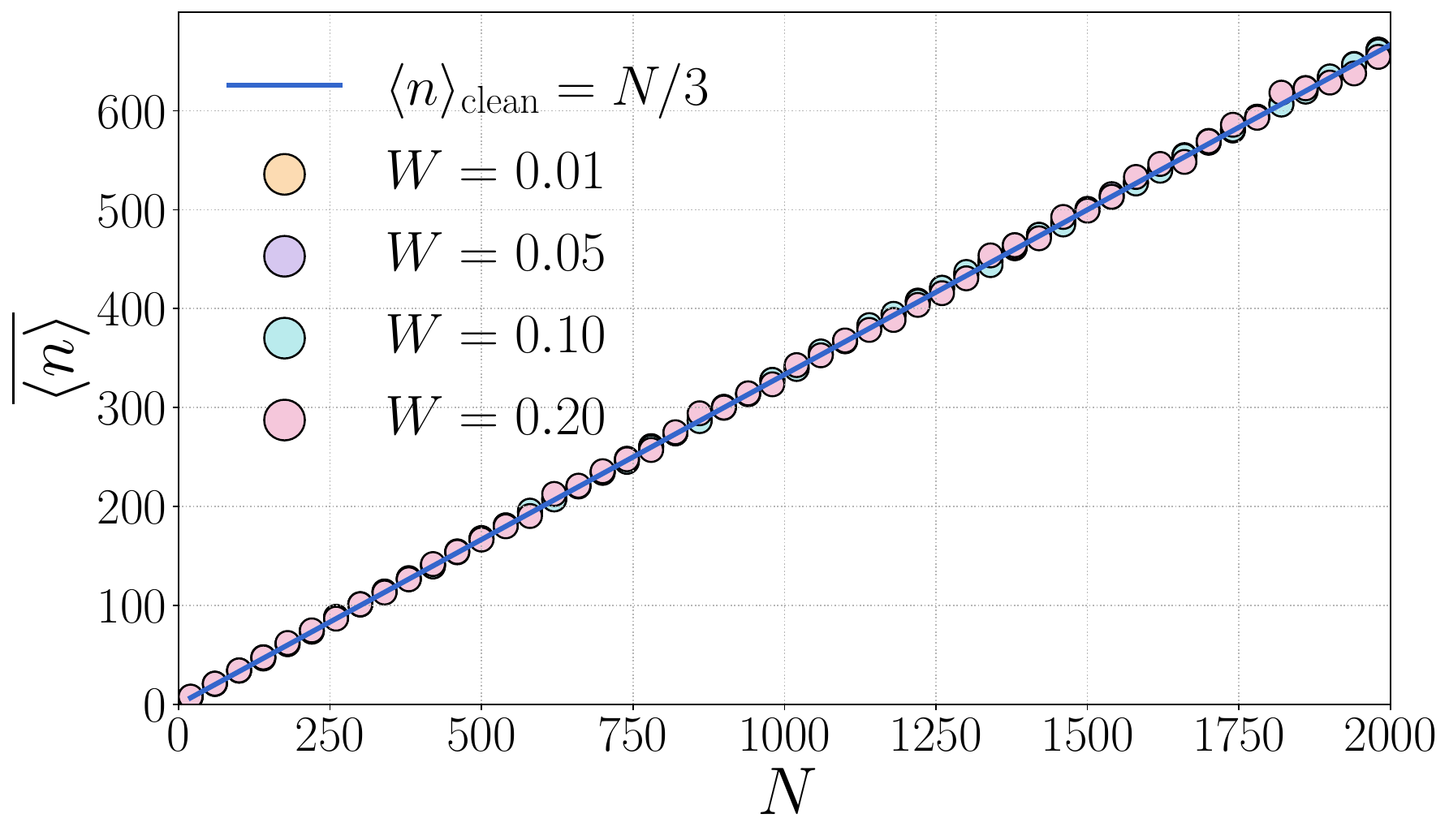}
    \includegraphics[width=0.48\linewidth]{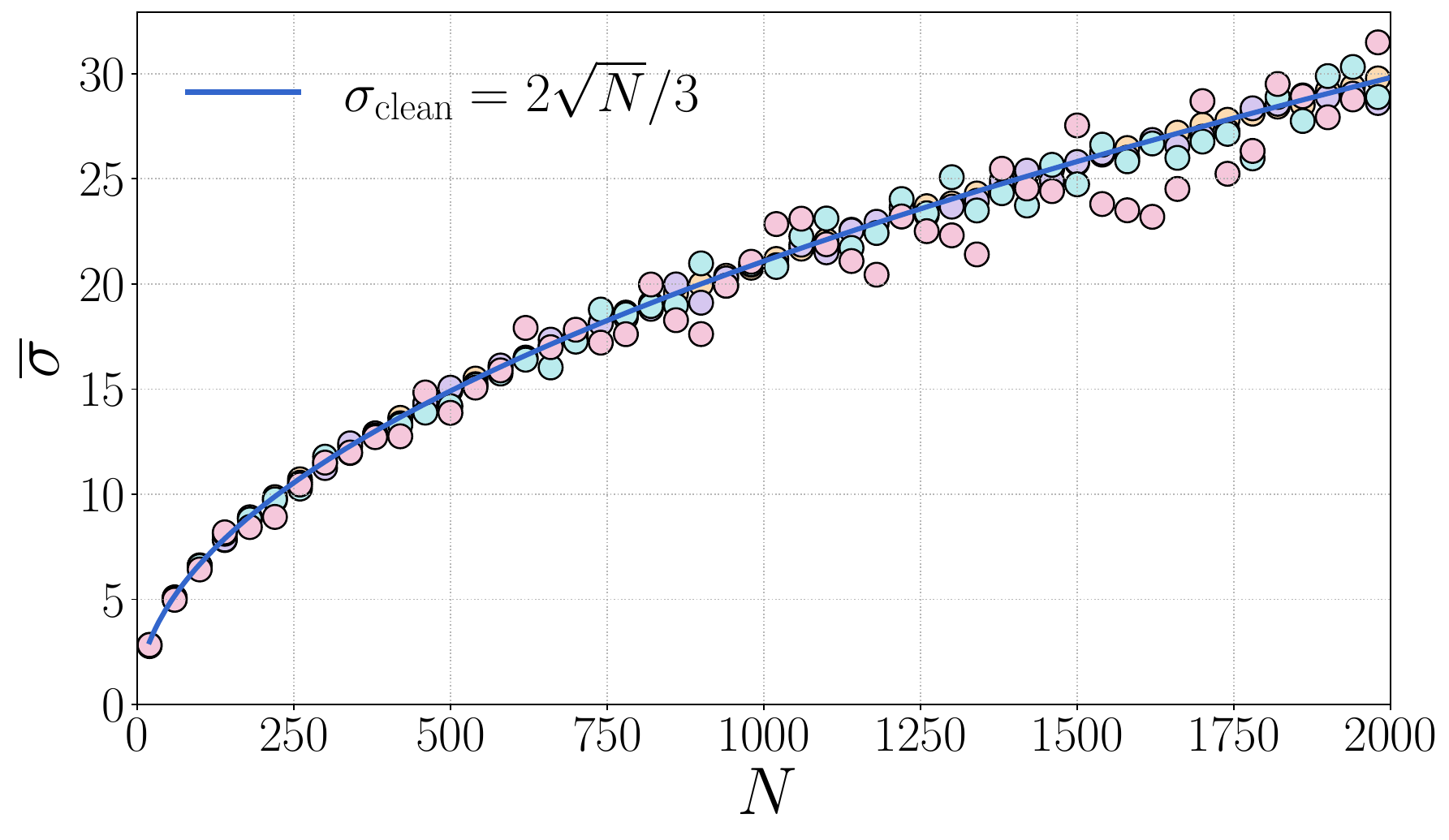}
    \caption{
    Disorder-averaged mean position $\overline{\langle n\rangle}$ (left) and root-mean-square width $\overline{\sigma}$ (right) of the zero mode for the power-law profile with $\alpha=1/4$ and $p=1/3$. The averages are taken over $20$ independent disorder realizations for various disorder strengths. The solid lines show the corresponding predictions for the clean model.
    }
    \label{fig:sigma_SM}
\end{figure}

To move beyond this qualitative observation, we characterize the zero mode by its mean position and root-mean-square width,
\begin{equation}
\langle n\rangle =\sum_n n\,|\Psi_n^{(0)}|^2, \qquad \sigma=\sqrt{\sum_n (n-\langle n\rangle)^2|\Psi_n^{(0)}|^2}.
\end{equation}
For the Gaussian profile predicted in the clean system, one has $\langle n\rangle=n_0=pN$ and $\sigma=\xi/\sqrt{2}$. The quantity $\sigma$ therefore provides a natural measure of the zero-mode extension even when weak disorder slightly distorts its profile. In Fig.~\ref{fig:sigma_SM}, we show the disorder-averaged quantities $\overline{\langle n\rangle}$ and $\overline{\sigma}$, obtained by averaging over $20$ independent disorder realizations, for several disorder strengths, using the power-law profile with $\alpha=1/4$ and $p=1/3$. The left panel shows that the average position of the zero mode remains equal to the clean interface position $n_0=pN$. The right panel shows that its width remains in excellent agreement with the clean prediction $\sigma=2\sqrt{N}/3$, indicating that the universal $\sqrt{N}$ scaling is preserved in the presence of weak disorder.

\end{document}